\documentclass[twocolumn]{aastex63}
\accepted{May 5, 2020}

\submitjournal{ApJL}

\shorttitle{CMB Shadows}
\shortauthors{Nashimoto et al.}

\begin{document}

\title{CMB Shadows: The Effect of  Interstellar Extinction on Cosmic Microwave Background Polarization and Temperature Anisotropy}

\correspondingauthor{Masashi Nashimoto}
\email{m.nashimoto@astr.tohoku.ac.jp}

\author[0000-0002-1221-1708]{Masashi Nashimoto}
\affiliation{Astronomical Institute, Tohoku University, Sendai, Miyagi 980-8578, Japan}
\affiliation{Graduate Program on Physics for the Universe (GP-PU), Tohoku University, Sendai, Miyagi 980-8578, Japan}
\author[0000-0003-0620-2554]{Makoto Hattori}
\affiliation{Astronomical Institute, Tohoku University, Sendai, Miyagi 980-8578, Japan}
\author[0000-0002-3266-857X]{Yuji Chinone}
\affiliation{Research Center for the Early Universe, School of Science, The University of Tokyo, Tokyo 113-0033, Japan}
\affiliation{Kavli Institute for the Physics and Mathematics of the Universe (Kavli IPMU, WPI), UTIAS, The University of Tokyo, Kashiwa, Chiba 277-8583, Japan}



\begin{abstract}
    We evaluate the degradation of the accuracy of the component separation between the cosmic microwave background (CMB) and foreground components caused by neglect of absorption of the monopole component of the CMB by the galactic interstellar matter. 
    The amplitude of the temperature anisotropy caused by the CMB
    shadow, due to dust components, is about 1\,$\mu$K. 
    This value is comparable to the required noise level necessary to probe non-Gaussianity studies with upcoming CMB experiments. 
    In addition, the amplitude of the polarization caused by the CMB shadow due to dust is comparable to or larger than the RMS value of the CMB $B$-mode polarization, imprinted by primordial gravitational waves. 
    We show that applying a single-power law model as the dust spectrum to observed multifrequency data introduces systematic errors, which are comparable to or larger than the required noise level for forthcoming CMB $B$-mode polarization experiments. 
    Deducing the intrinsic spectrum of dust emission from the submillimeter waveband data reduces systematic error below the required noise level. 
    However, this method requires dust temperature measurements with an accuracy of better than a few percent. 
    We conclude that the CMB shadow due to dust must be considered in future CMB missions for achieving their targeted sensitivity.
    Our results will be important to detect the primordial CMB $B$-mode polarization, with the amplitude of the tensor-to-scalar ratio of $r=10^{-3}$. 
\end{abstract}

\keywords{cosmic microwave background --- dust, extinction --- infrared: ISM --- submillimeter: ISM --- radiation mechanisms: general}

\section{Introduction} \label{sec:intro}
The galactic interstellar foregrounds are a serious obstacle for high-precision observations of the cosmic microwave background (CMB) temperature anisotropy and polarization. 
Separation of the CMB from synchrotron emission from relativistic electrons, free-free emission from ionized gases, and thermal emission from dust at microwave frequencies, has been widely studied \citep{radi-pro}. 
However, absorption of the CMB as a function of these components has not been considered in the component separation. 
According to Kirchhoff's law, emission mechanisms accompany finite absorption. 
The absorption of the CMB monopole due to the galactic interstellar matter causes apparent temperature anisotropy and $E$- and $B$-mode polarization.
In the present component separation methods, these effects are incorporated as part of the emission models. 
We refer to the apparent CMB temperature anisotropy and the CMB polarization imprinted by the absorption of the CMB monopole, due to the interstellar matter as the CMB shadow. 
The forthcoming CMB experiments will include LiteBIRD \citep{LiteBIRD_2014}; the CMB-S4 \citep{S4_2019}; the Simons Array \citep{SA_2014}; the Simons Observatory \citep{SO_2019}, requiring extremely high precision measurements of the CMB temperature anisotropy and the CMB polarization.
Improper treatment of the CMB shadow would have a significant impact on achieving these scientific goals.

In this paper, we perform quantitative studies of the CMB shadow and show that improper treatment of it could prevent upcoming CMB polarization experiments from realizing their full scientific potential.

\section{Apparent CMB temperature anisotropy due to interstellar absorption of the CMB monopole} \label{sec:int}
Apparent CMB temperature anisotropy, caused by the interstellar absorption of the CMB monopole, is explored.  
Main emission mechanisms of the galactic interstellar matter at microwave frequencies are the synchrotron emission, the free-free emission, and the emission from interstellar dust grains.
The absorption coefficient associated with synchrotron, which is defined as the optical depth per unit length, is given by Kirchhoff's law.
The absorption optical depth associated with the synchrotron emission, $ \tau^{\mathrm{sync}}_{\nu}$, due to relativistic electrons spiraling in the magnetic field, is is defined by the following equation (see, e.g., \citealt{radi-pro}),
\begin{eqnarray}
    I_\nu^\mathrm{sync} =
    \frac{\tau^\mathrm{sync}_\nu}{4\pi} 
    \frac{P^\mathrm{sync}_\nu}
    {\alpha_\nu^\mathrm{sync}},
    \label{eq:Inu_sync}
\end{eqnarray}
where $I_\nu^\mathrm{sync}$ is the observed intensity of the synchrotron emission, $P_\nu^\mathrm{sync}$ is the power of the synchrotron emission per unit volume per unit frequency, and $\alpha_\nu^\mathrm{sync}$ is the absorption coefficient associated with the synchrotron emission. 
The normalization of the intensity is chosen so that the brightness temperature at 408\,MHz is 20\,K \citep{PlanckX_2016}. 
This is a typical value in the galactic halo region. 
It is well-known that the single-power law model is not a good approximation of the spectrum of the galactic synchrotron emission from 400\,MHz through 100\,GHz.
The spectrum curves at approximately a few GHz \citep{Davies+2006}.
The main purpose of this paper is to estimate the order of magnitude of the synchrotron absorption: this includes modeling the energy distribution of relativistic electrons that emits synchrotron emission from 400\,MHz through 100\,GHz by a single-power law model, and to assess whether it is sufficient. 
The ratio $\alpha_\nu^\mathrm{sync}/P_\nu^\mathrm{sync}$ is calculated from the formulae described by \cite{radi-pro}. 
The synchrotron absorption optical depth is described by the following formula, 
\begin{eqnarray}
\tau^\mathrm{sync}_\nu =
\tau^\mathrm{sync}_0 
\left( \frac{\nu}{\nu_0} \right)
^{-\frac{p+4}{2}}, 
\label{eq:tau_sync}
\end{eqnarray}
where $p$ is the spectral index of the energy distribution of relativistic electrons, 
$\tau^\mathrm{sync}_0$ is the synchrotron absorption optical depth at $\nu_0=408$\,MHz, 
and 6\,$\mu$G is adopted as for the strength of the Galactic magnetic field 
component perpendicular to the line of sight \citep{Beck+2013}.

The absorption coefficient of free-free absorption is modeled as, 
\begin{eqnarray}
    \alpha_\nu^\mathrm{ff}
    =
    \alpha_0^\mathrm{ff}\, g_\mathrm{ff}(\nu,\, T_\mathrm{e})
    \left(\frac{\nu}{\nu_0}\right)^{-2},
    \label{eq:alpha_ff}
\end{eqnarray}
where $g_\mathrm{ff}$ is the Gaunt factor, which is a function of the frequency $\nu$ and the electron temperature $T_\mathrm{e}$.
We adopted the Gaunt factor model used in the \cite{PlanckXX_2011}.
The reference value $\alpha_0^\mathrm{ff}$ is set, so that EM is reproduced as 13\,cm$^{-6}$\,pc \citep{PlanckX_2016}.
This is a typical value in the galactic halo region.

The galactic dust emission is modeled as a superposition of thermal emission from dust grains, for which the frequency dependence of the emissivity is modeled by a single-power law model and the anomalous microwave emission (AME), originating from the spinning dust. 
Although the emission mechanism of AME has not been clarified yet (\citealt{Draine+1998}; \citeyear{Draine+1999}; \citealt{Nashimoto+2020}), quantitative differences of CMB absorption among these models are not significant. 
We adopt the spinning dust model prediction of the absorption coefficient, associated with AME, as a representative model. 

The absorption coefficient associated with the thermal emission from dust grains is described as,
\begin{eqnarray}
    \alpha_\nu^\mathrm{d}
    =
    \alpha_0^\mathrm{d} \left(\frac{\nu}{\nu_0}\right)^{\beta_\mathrm{d}}.
    \label{eq:alpha_d}
\end{eqnarray}
The reference value $\alpha_0^\mathrm{d}$ is determined, so that the optical depth of the dust is $4.50 \times 10^{-6}$ at 353\,GHz \citep{PlanckXI_2014}. 
This is a typical value in the galactic halo region.

The absorption coefficient due to spinning dust is expressed as \citep{Draine+2018}, 
\begin{eqnarray}
    \alpha_\nu^\mathrm{sp}
    &=&
    \alpha_0^\mathrm{sp} \left(\frac{\nu}{\nu_\mathrm{T}}\right)^4
    \exp\left[-\left(\frac{\nu}{\nu_\mathrm{T}}\right)^2\right],
    \label{eq:alpha_sp} \\
    \nu_\mathrm{T}
    &\equiv&
    \sqrt{\frac{15k_\mathrm{B}T_\mathrm{rot}}{16\pi^3\rho a^5}},
\end{eqnarray}
where $k_\mathrm{B}$ is the Boltzmann constant, $a$ is the dust radius, $\rho$ is the mass density of a dust grain, and $T_\mathrm{rot}$ is the rotational temperature.
Another reference value $\alpha_0^\mathrm{sp}$ is set, so that the peak value of spinning dust emission becomes $10^{-4}$ times the far infrared peak value of the dust thermal emission: this is the typical value in galactic clouds (e.g., \citealt{PlanckXV_2014}).
Applied values for each parameter are listed in Table \ref{tab:params}.
\begin{deluxetable}{lcc}
    \tablecaption{applied values for each parameter \label{tab:params}}
    \tablewidth{0pt}
    \tablehead{
    \colhead{Parameter} & \colhead{Value} & \colhead{Reference}}
    \startdata
    $p$ & 3 & --- \\
    $T_\mathrm{e}$ & 7000\,K & \cite{PlanckX_2016} \\
    $\beta_\mathrm{d}$ & 1.62 & \cite{PlanckXI_2014} \\
    $a$ & 5\,\AA  & --- \\
    $\rho$ & 2\,g\,cm$^{-3}$ & --- \\
    $T_\mathrm{rot}$ & 50\,K & \cite{Draine+2018} \\
    \enddata
\end{deluxetable}

The extinction of the CMB monopole due to scattering by the interstellar matter is negligibly small. 
First of all, the Thomson scattering of the CMB monopole by thermal electrons has no effect on the CMB intensity distribution and does not imprint any polarization signal.
Consider the scattering of the CMB monopole by a single electron. 
The electron scatters the CMB photon isotropically since the incident flux of the CMB monopole is isotropic and uniform. 
Therefore, no net effect on the intensity distribution is imprinted on the intensity distribution. 
Because there is no preferred direction for the electron, no polarized signal is imprinted. 
Spectrum distortion and polarized signal caused by the inverse Compton scattering of the CMB monopole by thermal electrons \citep{Birkinshaw_1999} are negligibly small because the Compton $y$-parameter is order of $10^{-10}$ and the optical depth times square of the ratio between tangential bulk velocity of the thermal electron system relative to the CMB rest frame (where we use a Solar System peculiar velocity of $369.0\, \mathrm{km\,s^{-1}}$ with respect to the CMB rest frame by \citealt{Hinshaw+2009}) and the speed of light is also order of $10^{-10}$. 
The number density of the relativistic electrons, which is evaluated by the energy equipartition between the energy density of the relativistic electrons with Lorentz factors from 300 to $10^4$ and that of the galactic magnetic field of 6\,$\mu$G, is eight orders of magnitude less than that of the thermal electrons.
Therefore, the scattering by the relativistic electrons is also negligible.
Rayleigh scattering due to dust in microwave is negligible because the scattering cross section is order of magnitude smaller than the absorption cross section. 

\begin{figure}[t!]
    \centering
    \includegraphics{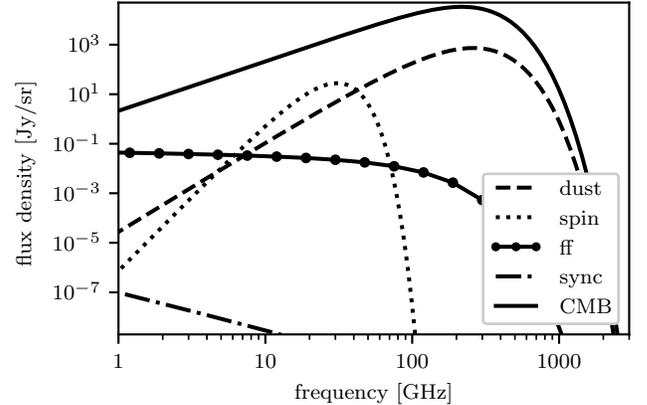}
    \caption{
    The spectra of the CMB shadows due to each component of the galactic interstellar matter.
    The solid curve is the spectrum of the CMB temperature anisotropy with $\delta T =70\,\mu$K.
    }
    \label{fig:cmbshadow_int}
\end{figure}
The spectra of CMB shadows due to each component of the galactic interstellar matter are shown in 
Fig.\,\ref{fig:cmbshadow_int}. 
For comparison, the spectrum of the CMB temperature anisotropy with $\delta T =70\,\mu$K, which is the RMS value at $\ell\simeq180$, is overlaid.
The amplitude of the temperature anisotropy caused by the CMB shadow due to dust is about 1\,$\mu$K. 
This is approximately 1\% of the RMS value of the first acoustic peak of the CMB temperature anisotropy, and is comparable to the required noise level to achieve the goal of non-Gaussianity studies by next-generation CMB experiments \citep{Sohn+2019}.
Their relative contributions to the CMB temperature anisotropy become smaller as the frequency becomes lower from their peak frequencies.
The synchrotron absorption and free-free absorption is less than 0.1\,$\mu$K in the frequency band, higher than 10\,GHz.
Therefore, the CMB shadow has a negligible contribution for extracting the first acoustic peak of the CMB temperature anisotropy from the microwave data in the galactic halo region. 
However, we must pay attention to the CMB shadow when the component separation toward the galactic disc directions and molecular clouds are carried out. 
Further, the CMB shadow presents a nonnegligible effect in order to achieve the goal of non-Gaussianity studies, based on next-generation CMB experiments \citep{Sohn+2019}.

\section{Polarization caused by interstellar absorption of the CMB monopole} \label{sec:pol}
The polarization component of the CMB shadow, due to each interstellar component, is estimated by multiplying Eqs.\,(\ref{eq:tau_sync})--(\ref{eq:alpha_sp}) by the degree of polarization.
For simplicity, the frequency dependence of the degree of polarization is neglected in this study.
The polarization degree of synchrotron emission is set at 10\%, which is 3\%--5\% toward the galactic plane and increases above 20\% with increasing the galactic latitude \citep{Kogut+2007}.
Since the free-free emission is not polarized except at the edges of the H\,$_\mathrm{II}$ regions, the polarization degree of the free-free emission is assumed to be zero \citep{Macellari+2011}.
The degree of polarization of the thermal emission from dust is set at 7\% \citep{PlanckXIX_2015}.
Polarized AME has not been detected so far, and it is unclear if AME is polarized.
The deepest upper limit, regarding polarization degree of the AME, is given by QUIJOTE as 0.1\%--1\% (\citealt{QUIJOTE1_2015}; \citeyear{QUIJOTE2_2017}).
Prediction of the polarization degree of the AME is ranging from a few percent \citep{Draine+2013} to $10^{-4}$\,\% \citep{Draine+2016}. 
Considering the number of model predictions of the polarization degree, we examine two cases. 
One occurs when the polarization degree of the AME equals the observed upper limit of 0.1\%.
The other occurs when the polarization degree of the AME equals the spinning dust model prediction of $10^{-4}\,$\%.
Since it is observationally clear that one-third of the polarization spectra of synchrotron emission and dust thermal emission contribute to the $B$-mode (\citealt{SPASS_2018}; \citealt{PlanckXXX_2016}), one-third of the degree of polarization is considered as the degree of polarization for the $B$-mode polarization emission. 
Therefore, one-third of the polarization of CMB shadows are treated as their $B$-mode components. 
The $E$-mode of the CMB shadow is twice as strong as the $B$-mode of the CMB shadow.

\begin{figure}[t!]
    \centering
    \includegraphics{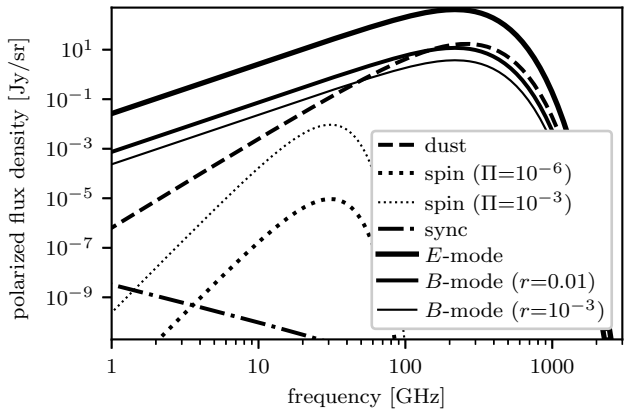}
    \caption{
        The polarized spectra of the CMB shadows due to each component of the galactic interstellar matter.
    }
    \label{fig:cmbshadow_pol}
\end{figure}
Fig.\,\ref{fig:cmbshadow_pol} shows the spectra of the $B$-mode polarization of the CMB shadow due to each component of the galactic interstellar matter. 
The spectrum of the CMB $E$-mode polarization, with the amplitude of the RMS value, calculated the best-fitting $\Lambda$CDM model from \textit{Planck}, using \textsf{CAMB} \citep{Lewis+2000} at around $\ell=100$, or 0.86\,$\mu$K, is overlaid. 
Since the amplitude of the $E$-mode polarization originating from the CMB shadow due to dust is twice the amplitude of the $B$-mode due to the CMB shadow \citep{PlanckXXX_2016}, the amplitude of the $E$-mode polarization caused by the CMB shadow due to dust is estimated at about 0.02\,$\mu$K at the peak frequency, as seen in Fig.\,\ref{fig:cmbshadow_pol}. 
The high precision measurement of the reionization bump appearing in the CMB $E$-mode polarization power spectrum requires a noise level of $E$-mode polarization measurement, better than 0.03\,$\mu$K (\citealt{LiteBIRD_2014}; \citealt{SO_2019}; \citealt{S4_2019}). 
The CMB shadow introduces systematics at the required noise level.

In Fig.\,\ref{fig:cmbshadow_pol}, the CMB $B$-mode polarization spectra are overlaid. 
Their amplitudes of the RMS values are predicted from the value of the CMB $B$-mode polarization at the recombination bump, originating from the primordial gravitational waves, with an amplitude of $r=0.01$ and $r=10^{-3}$, which are targeted by the current CMB experiments and the upcoming CMB experiments, respectively.
It shows that the CMB $B$-mode polarization, due to the absorption associated with the synchrotron emission, is negligibly small compared to that of the $B$-mode polarization in the frequency range higher than 10\,GHz. 
On the other hand, the amplitude of the $B$-mode polarization caused by the CMB shadow due to the dust exceeds the amplitude of the CMB $B$-mode polarization with $r=0.01$ at frequencies above 100\,GHz.

\section{Systematic errors introduced by improper treatment of the CMB shadow}\label{sec:discuss}
In this section, we quantify the error made by neglect of CMB monopole absorption when modeling thermal dust emission with a power-law emissivity.
Since the effect is much more significant to the CMB $B$-mode polarization measurement than the $E$-mode measurement, the quantitative estimation is performed for the $B$-mode measurement. 
The effective dust polarization spectrum $P_\nu^\mathrm{eff}$ after subtracting the CMB shadow is described by:
\begin{eqnarray}
    P_\nu^\mathrm{eff}
    &=&
    P_\nu^\mathrm{d}
    - \tau_\nu^\mathrm{d} \Pi_\nu^\mathrm{d} B_\nu(T_\mathrm{CMB})
    \nonumber \\
    &=&
    \tau_\nu^\mathrm{d} \Pi_\nu^\mathrm{d}
    \left[B_\nu(T_\mathrm{d}) - B_\nu(T_\mathrm{CMB}) \right],
    \label{eq:Peff}
\end{eqnarray}
where $T_\mathrm{d}$ is the dust temperature, which is about 19.7\,K \citep{PlanckXI_2014}, $T_\mathrm{CMB}$ is the CMB temperature, which is 2.725\,K \citep{Mather+1999}, $\tau_\nu^\mathrm{d}$ is the dust optical depth, $\Pi_\nu^\mathrm{d}$ is the polarization fraction of dust, $P_\nu^\mathrm{d}$ is the intrinsic dust polarization spectrum, and $B_\nu$ is the Planck function. 
Since the direction of the polarization caused by the CMB shadow due to dust is perpendicular to the polarization direction of the dust emission, the polarization caused by the CMB shadow is subtracted from the polarization intensity of the dust emission in Eq.\,(\ref{eq:Peff}).
In the current standard polarization experiment, the dust polarization spectrum is modeled as a single-power law spectrum. 
It is self-evident that the effective dust polarization spectrum is not described by a single-power law model in any frequency range, even if the intrinsic spectrum of thermal emission from the dust is adequately approximated by a single power-law model.

\begin{figure}[t!]
    \centering
    \includegraphics{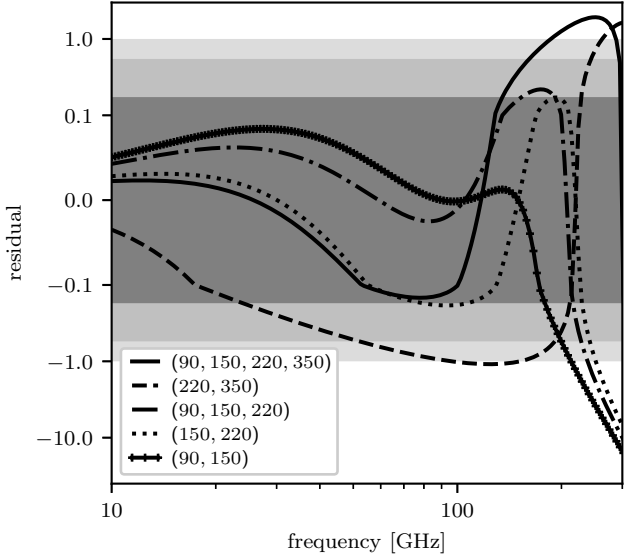}
    \caption{
    The difference in the effective polarization emission from dust estimated by the single-power law fitting for the effective polarization emission.
    The line styles of the curves correspond to the frequency bands used for fitting.
    The grey colored regions correspond to $r < 3\times 10^{-4}$, $3\times 10^{-3}$, and 0.01 in order of darkness.
    }
    \label{fig:residual}
\end{figure}
The systematic errors introduced by applying a single-power law model as dust spectra are quantitatively estimated. 
The intensity spectrum of the thermal emission from the dust is assumed to be described by a single-power law model, with a spectral index of $\beta_\mathrm{d}=1.62$. 
The contribution of the CMB $B$-mode polarization itself is neglected. 
We suppose that noiseless polarization observations are performed at 90, 150, 220, and 350\,GHz. 
The delta function is adopted as a bandpass model for simplicity.
The degree of polarization is assumed to be constant over this frequency range. 
Five combinations of the observed frequency bands are considered. 
For each case, the best-fit single-power law model is obtained by least-square fitting to the observed distribution of the  polarization intensities.  
Fig.\,\ref{fig:residual} shows residuals after subtracting the best-fit single-power law model prediction from the true effective dust $B$-mode polarization spectrum.  
The amplitude of the residual is normalized by the $B$-mode polarization spectrum with the amplitude of $r=0.01$.
If the contribution from the CMB shadow is zero, the amplitude of the residual is also zero.
It shows that when two frequency bands are used to subtract the dust contribution with a single-power law model, selecting 90 and 150\,GHz is the best choice, as this combination has the smallest residual around the CMB peak frequency. 
The inclusion of the higher frequency band increases the residual, and renders less precision in the accuracy of the fitting.
When the effective dust polarization spectrum is fitted by a power law using only high frequency bands (220 and 350\,GHz), the residual is large and comparable to the amplitude of the $B$-mode polarization spectrum with $r = 0.01$ from 100 to 160\,GHz. 
Except for the combination of 90 and 150\,GHz bands, the residual is comparable to or larger than $\sigma_r=3\times10^{-4}$ in the frequency range of 90 to 150\,GHz. 
The 1-$\sigma$ error of the tensor-to-scalar ratio of $\sigma_r=3\times10^{-4}$ is the requirement to achieve 3-$\sigma$ detection of the CMB $B$-mode polarization with the amplitude of $r=10^{-3}$ \citep{LiteBIRD_2014}.

\begin{figure}[t!]
    \centering
    \includegraphics{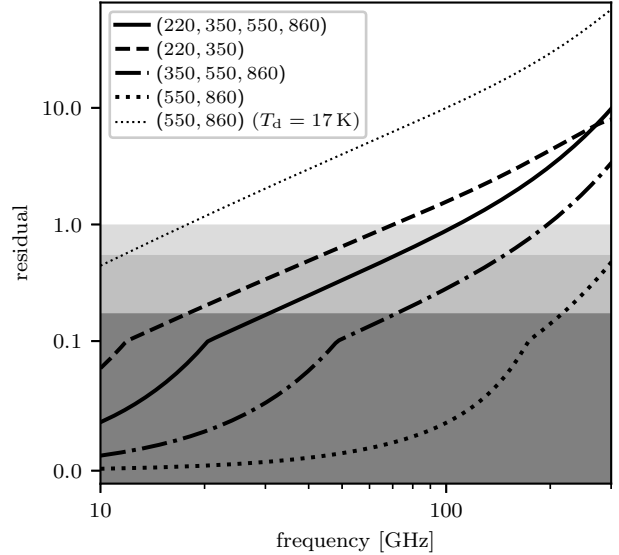}
    \caption{
    The difference in the effective polarization spectrum from dust calculated by the power-law fitting for the dust optical depth.
    The line styles of the curves correspond to the frequency bands used for fitting.
    The thin dotted curve is the residual when the dust temperature is misestimated as $T_\mathrm{d} = 17$\,K.
    The grey colored regions correspond to $r < 3\times 10^{-4}$, $3\times 10^{-3}$, and 0.01 in order of darkness.
    }
    \label{fig:residual_def}
\end{figure}
One solution to overcome the degradation of the single-power law model fitting, in multifrequency band observations, is proposed as follows. 
First, the dust polarization spectrum in the submillimeter wavebands is fitted by a single-power law model to extract the intrinsic dust polarization spectrum. 
Next, the effective dust polarization spectrum in the CMB frequency bands is estimated by using Eq.\,(\ref{eq:Peff}). 
Fig.\,\ref{fig:residual_def} shows the residual after subtracting the dust polarization spectrum, obtained by applying this method
from the true effective dust polarization spectrum. 
Four combinations of observed frequency bands to deduce the dust polarization spectrum are considered, as seen in Fig.\,\ref{fig:residual_def}, showing that the combination of 550 and 860\,GHz is the best choice to minimize the systematics.
The inclusion of 350\,GHz degrades the fitting results drastically, because the CMB shadow has a nonnegligible effect at 350\,GHz. 
Even with the combination of 550 and 860\,GHz bands, systematic errors can be significant if dust temperature is not correctly estimated.
In Fig.\,\ref{fig:residual_def}, the residual when the dust temperature is misestimated as $T_\mathrm{d}=17$\,K, is also overlaid.
This demonstrates that uncertainty in 10\% of dust temperature measurements prevents achieving the $B$-mode polarization detection with $r=0.01$. 
The dust temperature must be measured to an accuracy of a few percent.
Accurate estimation of dust temperature by observations of far-infrared peaks is important in reducing systematic errors due to the CMB shadow.
It is interesting to study how the accuracy of dust temperature measurement is improved by using far-infrared all-sky diffuse maps provided by an astronomical infrared satellite, e.g. AKARI \citep{Doi+2015}.

\section{Summary}\label{sec:summary}
The effect of the absorption of the monopole component of the CMB by galactic interstellar matter, on the degradation of the accuracy of the component separation between the CMB and the foreground components, was evaluated for the first time. 
The amplitude of the temperature anisotropy caused by the absorption of the CMB monopole, due to interstellar matter, is superposed on foreground emission as negative emission, both in intensity and polarization.
The CMB shadow due to galactic dust has a nonnegligible effect on the high precision measurement. 
The amplitudes of the temperature anisotropy, caused by the CMB shadow due to dust components, are about 1\,$\mu$K. 
This is about 1\% of the RMS value of the first acoustic peak and comparable to the required noise level to achieve the goal of non-Gaussianity studies by the next-generation CMB experiments.  
The amplitude of the polarization caused by the CMB shadow due to dust is comparable to or larger than the RMS value of the CMB $B$-mode polarization imprinted by primordial gravitational waves.
We show that applying a single-power law model to fit observed multifrequency dust spectrum data introduces systematic errors, comparable to or larger than the required noise level for the forthcoming CMB $B$-mode polarization experiments.
Deducing the intrinsic spectrum of the dust emission, using data for submillimeter wavebands, could reduce the systematic error below the required noise level. 
This method requires the dust temperature measurements with an accuracy of higher than a few percent. 
We conclude that the CMB shadow due to dust must be considered for future CMB missions to achieve their targeted sensitivity in detecting of CMB $B$-mode polarization.

\acknowledgments
MN acknowledges support from the Graduate Program on Physics for the Universe (GP-PU), Tohoku University.
YC acknowledges the support from the JSPS KAKENHI Grant Number 18K13558.
This work is partially supported by MEXT KAKENHI Grant Number 18H05539.


\bibliography{article}{}
\bibliographystyle{aasjournal}

\end{document}